\documentclass[reprint,amsmath,amssymb,aps]{revtex4-2}

\usepackage{graphicx}
\usepackage{dcolumn}
\usepackage{bm}
\usepackage{acronym}
\usepackage{color}
\begin{document}
\preprint{APS/123-QED}
\title{Cosmological Inference using Gravitational Waves and Normalising Flows}

\author{Federico Stachurski}
\email{f.stachurski.1@research.gla.ac.uk}
\author{Christopher Messenger}
\author{Martin Hendry}

\affiliation{SUPA, School of Physics and Astronomy, University of Glasgow, Glasgow, United Kingdom}

\begin{abstract}
We present a machine learning approach using normalising flows for inferring cosmological parameters from gravitational wave events. Our methodology is general to any type of compact binary coalescence event and cosmological model and relies on the generation of training data representing distributions of gravitational wave event parameters. These parameters are  conditional on the underlying cosmology and incorporate prior information from galaxy catalogues. We provide an example analysis inferring the Hubble constant using binary black holes detected during the O1, O2, and O3 observational runs conducted by the advanced LIGO/VIRGO gravitational wave detectors. We obtain a Bayesian posterior on the Hubble constant from which we derive an estimate and 1$\sigma$ confidence bounds of $H_{0} = 74.51^{+14.80}_{-13.63} \: \text{km} \:\text{s}^{-1} \text{Mpc}^{-1}$. We are able to compute this result in $\mathcal{O}(1)$ s using our trained Normalising Flow model. 
\end{abstract}

\maketitle
\acrodef{GW}[GW]{gravitational wave}
\acrodef{CBC}[CBC]{compact binary coalescence}
\acrodef{NF}[NF]{normalising flow}
\acrodef{BBH}[BBH]{binary black hole}
\acrodef{SNR}[SNR]{signal-to-noise ratio}
\acrodef{NSBH}[NSBH]{neutron star black hole}
\acrodef{EM}[EM]{electromagnetic}
\acrodef{MLP}[MLP]{Multi Layer Perceptron}
\acrodef{PE}[PE]{parameter estimation}
\acrodef{PSD}[PSD]{power spectral density}
\acrodef{GPU}[GPU]{graphics processing unit}

\emph{Introduction.}---  The idea of using \ac{GW} detections to gain insights into the cosmological properties of our universe was first proposed by Schutz~\cite{Schutz}. These "standard sirens" give us information about the calibrated luminosity distance of the event without the use of the cosmological distance ladder or any prior knowledge or assumptions of the universe. By adding the information of redshifts from galaxy catalogs and using the relationship between luminosity distance and redshift~\cite{Hogg} one can infer cosmological parameters.

At present, the measurements of $H_{0}$ are in tension with each other, of approximately 4.4$\sigma$ \cite{freedman2017cosmology}. The Planck experiment~\cite{Planck_2020} estimated the Hubble constant to be $H_{0} = 67.4 \pm 0.5$ km s$^{-1}$Mpc$^{-1}$ ($1\sigma$  confidence interval), using measurements from the cosmic microwave background radiation. The SH0ES experiment \cite{Riess_2019} measured $H_{0} = 74.03 \pm 1.42$  km s$^{-1}$Mpc$^{-1}$ ($1\sigma$ confidence interval), making use of the measured distances of Type 1a supernova standard candles. This suggests that either one of the experiments might be subject to unknown systematic errors, or maybe an indication of some "new" underlying physics causing this discrepancy~\cite{freedman2017cosmology}. For this reason, \ac{GW} standard siren measurements of $H_{0}$ will be beneficial, hopefully, in breaking this tension between the current values. 
Since the LIGO and Virgo detectors ~\cite{Abbott_2020,LIGOScientific:2014pky,VIRGO:2014yos} were activated, they have revealed a significant number of \ac{GW}s events—totaling 90 thus far. These events span a range of phenomena, from \acp{BBH} mergers to  \acp{NSBH}, as well as a binary neutron star (BNS) coalescences~\cite{ligo_gwtc3, Pop_paper_2021}.

The most recent results of the Hubble constant estimation using \ac{GW}s are from \cite{COSMO_2022}, where 2 independent analysis methods were employed. In the first, \texttt{gwcosmo}~\cite{PhysRevD.101.122001}, they provide an estimated value of $H_{0} = 68 ^{+12}_{-8} \: \text{km} \:\text{s}^{-1} \text{Mpc}^{-1}$ ($1\sigma$ highest density interval) when combined with the $H_{0}$ measurement from GW170817 and its electromagnetic radiation counterpart. 
In the second analysis~\cite{ICAROGW}, no galaxy catalog prior information was used. We also note Hubble value estimates of $H_{0} = 68 ^{+26.0}_{-6.2} \: \text{km} \:\text{s}^{-1} \text{Mpc}^{-1}$ ($1\sigma$ equal-tailed interval) obtained using spatial cross-correlation between \ac{GW} sources and the photometric galaxy surveys~\cite{Mukherjee:2022afz, Mukherjee:2020hyn, Diaz:2021pem}.


In this work we will see how our new analysis, \texttt{CosmoFlow}, using \acp{NF}, a machine learning driven process which allows us to define expressive probability distributions~\cite{NF}, over cosmological parameters using \ac{GW} posterior samples as inputs. \acp{NF} are widely used in the \ac{GW} community, from performing fast and reliable \ac{GW} parameter estimation~\cite{Williams_2021, williams2023importance, DINGO} to population studies \cite{ruhe2022normalizing}. In this work we demonstrate that a \ac{NF} can model the \acf{EM} dependent prior of the \ac{GW} parameters. We can then account for selection effects and uncertainties in the detected \ac{GW} parameters to evaluate the likelihood of a cosmological parameter set. We train a \ac{NF} model prior to the detection of \ac{GW} events and for each event we input parameter estimation results and compute the cosmological parameter likelihood. The time taken by the \ac{NF} is $\sim 1$ ms per set of cosmological parameters. This allows us to compare our analysis to that of \texttt{gwcosmo}~\cite{COSMO_2022} and compute an overall combined posterior distribution over the Hubble constant with the 42 \acp{BBH} observed during the O1, O2 and O3 observing runs in the order of $\mathcal{O}(1)$ s.

\emph{Bayesian Framework.}--- We define the posterior distribution of the cosmological parameters, $\Omega$, conditioned by the \ac{GW} strain data of an ensemble of $n$ events, $\mathbf{h}=[h_1,h_2,\ldots,h_n]$ as 
\begin{align}\label{eq:posterior}
    p(\Omega|\mathbf{h},\mathbf{D},I) = p(\Omega|I)\prod_i \frac{p(h_i,D_i|\Omega,I)}{p(h_i,D_i|I)},
\end{align}
where $\mathbf{D} = [D_1,D_2,\ldots,D_n]$ is the binary state of detection, 1 for detected and 0 for not detected for the $i$th event, and $I$ represents all other assumed information. The detectability of a \ac{GW} signal, serves to account for the selection effects that arise when applying \ac{SNR} thresholds on candidate events and we must make sure that the likelihood term is properly normalised, such that $\sum_{D_i=0,1}\int p(h_{i},D_{i}|\Omega,I) dh_i = 1$. However, since encoded within $I$ is the information that only events exhibiting an \ac{SNR} greater than some threshold $\rho_{\text{th}}$ are considered, we find that 
\begin{align}\label{eq:liknorm}
p(h_i,D_i|\Omega,I) = \frac{p(h_i|\Omega,I)}{p(D_i|\Omega,I)}
\end{align}
since by definition, $p(D_i=0|\Omega,I)=0$ for all detected events. This result allows us to write our cosmological parameter posterior as 
\begin{align}\label{eq:posterior_2}
    p(\Omega|\mathbf{h},\mathbf{D},I) = p(\Omega|I)\prod_i \frac{\int p(h_{i}|\theta_i,I)p(\theta_i|\Omega,I)\,d\theta_i}{p(h_i,D_i|I)p(D_{i}|\Omega,I)}.
\end{align}
where we have marginalized numerator of Eq.~\eqref{eq:liknorm} over the \ac{GW} parameters, $\theta$.

It will become apparent when we discuss the generation of training data for our \ac{NF} approach that it is more practical to deal with a \ac{GW} parameter prior that is conditional on detection. Hence via Bayes' theorem we obtain 
\begin{align}\label{eq:detflip}
p(\theta_i|\Omega,D_i,I) = \frac{p(D_i|\theta_i,I)p(\theta_i|\Omega,I)}{p(D_i|\Omega,I)}
\end{align}
where we have used the fact that the detectability of an event is independent of $\Omega$ if the \ac{GW} parameters are given.

With some rearrangement and using Eq.~\eqref{eq:detflip} and replacing the \ac{GW} likelihood with the ratio of its posterior and prior, we are able to write our cosmological parameter posterior as
\begin{align}\label{eq:posterior_3}
    p(\Omega|\mathbf{h},\mathbf{D},I) \propto p(\Omega|I)\prod_i \int\frac{ p(\theta_i|h_{i},I)p(\theta_i|D_i,\Omega,I)}{p(D_i|\theta_i,I)p(\theta_i|\Omega_0,I)}\,d\theta_i,
\end{align}
where $p(\theta|\Omega_0,I)$, represents the \ac{GW} parameter priors used in the parameter estimation assuming a fixed cosmology $\Omega_0$~\cite{Pop_paper_2021}.

We can then approximate the integral over the \ac{GW} parameters as a Monte Carlo summation giving us  
\begin{align}\label{eq:master1}
    p(\Omega|\mathbf{h},\mathbf{D},I) \propto p(\Omega|I)\prod_i\Bigg \langle \frac{p(\theta_i|D_i,\Omega,I)}{p(D_i|\theta,I)p(\theta|\Omega_0,I)} \Bigg \rangle_{\theta \sim p(\theta_i|h_i,I)},
\end{align}
as our final result.

The numerator within Eq.\eqref{eq:master1} represents the \ac{GW} parameter priors conditional on detection and the cosmological parameters. We will use a \ac{NF} to model this prior and incorporate \ac{EM} galaxy catalogue information within the training procedure. For the two terms in the denominator, we use the survival function of a non central $\chi$-squared distribution with non central parameter the \ac{SNR} squared and d.o.f. $k = 2n$, where $n$ is the number of detectors used for the detection of the event, to compute $p(D_i|\theta,I)$, instead the term $p(\theta|\Omega_{0},I)$ is the prior on the GW parameters used in the \acp{PE} process conditioned by a fixed cosmology \cite{Pop_paper_2021}.

\emph{Normalising Flows.}--- \acp{NF} have the capacity to efficiently evaluate and sample from complex probability distribution functions. They operate by transforming a simple data distribution, such as a multivariate Normal distribution, through a series of affine transformations, ultimately generating a more intricate output distribution~\cite{NF, NF_papa}. 
Assuming $x$ is a random variable sampled from a distribution $p_{x}(x|\omega)$, and that another random variable, $y$, sampled from distribution $y \sim p_{y}(y|\omega)$ is related with $x$ following the relation $y = g(x)$ and $x = f(y)$, where $\omega$ is a conditional statement, then a flow can be constructed with a conditional statement \cite{papamakarios2018masked}, which ultimately allows for the computation of likelihood-like terms, such as the one in Eq.\eqref{eq:master1} \cite{conditionalflow}. Therefore, using the change of variable equation becomes:

\begin{multline}\label{flow_eq}
    \log(p_{\mathbf{x}}(\mathbf{x}|\omega)) = \log(p_{\mathbf{y}}(f^{-1}(\mathbf{x}, \theta|\omega))) + \\ \log\bigg(\text{det}\bigg|\frac{df^{-1}(\mathbf{x}, \theta|\omega)}{d\mathbf{x}}\bigg| \bigg) , 
\end{multline}
where $\theta$ are trainable parameters for the function $f$. The $\theta$ parameters are optimised with a loss function, using the Kullback–Leibler divergence between the left hand side and the right hand side of Eq.\eqref{flow_eq}.

Therefore, a set of non-linear functions can be composed together to construct more complicated functions, allowing to go from a simple distribution to a more complicated data distribution \cite{NF}. Using Eq.\eqref{flow_eq}, it is possible to construct a flow to compute the numerator term in Eq.\eqref{eq:master1}, where $x$ are the \ac{GW} parameters and $\omega = [H_{0}, D_{i}]$.

\emph{Data Generation.}--- In order to train our \ac{NF} we must provide it with training data representing the \ac{GW} parameter distribution we wish to model. We must also provide the  corresponding cosmological parameters upon which that distribution is conditioned. The data generation process and results presented in this manuscript have been limited to the case where inference is performed only on the Hubble constant. We have restricted the analysis for the sole purpose of direct validation of our results with those presented in~\cite{COSMO_2022}.

We begin by sampling cosmological values $\Omega$, where in this case $\Omega\equiv H_0$, which will serve as the conditional input for the \ac{NF}. We consider a specific cosmology characterized by a Flat$\Lambda$CDM model with fixed cosmological parameters $\Omega_{m} = 0.3$ and $w_{0} = -1$, as used in~\cite{Gray_2020, COSMO_2022}, and vary the expansion rate parameter $H_{0}$ uniformly within the range $[20,140]\,\text{km}\,\text{s}^{-1}\text{Mps}^{-1}$.
 
We use the sampled values of $H_{0}$ to compute the Schechter functions $\phi(L|\Omega)$ from which we sample the luminosities  of host galaxies. The Schechter functions are those described in the GLADE+ catalog~\cite{GLADE}, using the $K$-band fitting parameters given in \cite{COSMO_2022}.  
We perform \emph{k}-corrections and color-evolution corrections~\cite{COSMO_2022} and introduce luminosity weighting, favouring galaxies in proportion to their luminosities as hosting \ac{GW} events~\cite{Gray_2020}, hence $p(L|\Omega)\propto L\phi(L|\Omega)$.

We next sample the redshifts of galaxies hosting \ac{BBH} events from a prior uniform in co-moving volume and co-moving time, multiplied by the merger rate component, described in Appendix A.2 in~\cite{COSMO_2022}, where the fitting parameters from~\cite{Pop_paper_2021} were used~\footnote{In general this distribution is dependent on the cosmological parameters $\Omega$ with the exception of the Hubble constant.}.
The redshifts are then combined with their corresponding cosmological parameters to obtain the luminosity distances of galaxies hosting \ac{GW} events. With the sampled galaxy luminosity we can then compute the apparent magnitude of the galaxy.

At this point we sample uniformly over the 2-sphere to generate sky locations. 
To incorporate redshift and sky location information from the local universe into our training data, we use the GLADE+~\cite{GLADE} catalog. To select galaxies from the catalog, we compare the apparent magnitude and location of the sampled galaxy with the magnitude threshold map of the sky~\cite{GrayTh}, we determine if such a galaxy would have been contained within the catalog for the corresponding sky pixel location. If false, we retain the sampled galaxy attributes and record that the host is outwith the catalog. However, if true, we substitute the previously sampled galaxy with a randomly sampled galaxy (weighted by $(1+z)^{-1}$ and in proportion with luminosity) from that pixel in the catalog. The sample will retain its sampled cosmological parameter but apparent magnitude and sky location will be taken from the selected catalog galaxy. The catalog redshift is interpreted as an uncertain measurement of the redshift which is generated by sampling from a Gaussian distribution centred on the measured value with standard deviation taken from the catalog redshift uncertainty. This redshift is combined with the cosmological parameter to compute the luminosity distance. 

With host galaxy parameters selected, we proceed to sample the \ac{GW} parameters of the event. The parameters we sample are: $m_{1}$ and $m_{2}$, primary and secondary masses, using a power-law plus peak model for $m_{1}$ and a power-law for $m_{2}$ conditioned on $m_{1}$~\cite{Pop_paper_2021, COSMO_2022}; $a_{1}$ and $a_{2}$, primary and secondary spins, from a uniform distribution between $[0, 0.99]$ for both parameters~\cite{ligo_gwtc3}; $\theta_{1}$, $\theta_{2}$ and $\theta_{JN}$, primary and secondary axis orientation and the orbital plane inclination angle, respectively, from uniform distributions between $[-1, 1]$ in $\cos(\theta)$~\cite{ligo_gwtc3}; $\phi_{JL}$,  $\phi_{12}$ and $\psi$, from a uniform distributions between $[0, 2\pi]$~\cite{ligo_gwtc3}. Lastly we sample geocentric time of arrival from a uniform distribution over one sidereal day, $[0.0000, 86164.0905] \:  \text{s}$. This comprises our set of 11 intrinsic \ac{GW} parameters (omitting a reference phase) plus the extrinsic sky position and luminosity distance parameters shared with the host galaxy. 

After sampling \ac{GW} parameters, we employ the \texttt{BILBY} package~\cite{Bilby} to simulate event-specific \ac{SNR} to determine whether an event would be detected. We produce datasets that correspond to various detector configurations, considering the relevant \acp{PSD} employed during the O1, O2, O3a, and O3b observational periods. Each data set is then used to train an individual flow for that specific detector setup and observational period. 

Under the assumption of well behaved Gaussian detector noise we draw samples of matched-filter \ac{SNR} from a non-central $\chi$ distribution with non-centrality parameter equal to the optimal network \ac{SNR} and with $2n$ degrees of freedom where $n$ is the number of detectors in the network. If the sampled \ac{SNR} is $\geq$ than \ac{SNR} threshold ($\rho_{\text{th}}=11$ ~\cite{COSMO_2022}) then the event is retained, otherwise the sample is discarded and we begin the sampling procedure again from the start but retaining the original sample of the cosmological parameters. This latter choice ensures that all sampled parameters will be conditional upon detection with the exception of our cosmological parameters.

\begin{figure*}[t]
    \includegraphics[ width=1.0\textwidth]{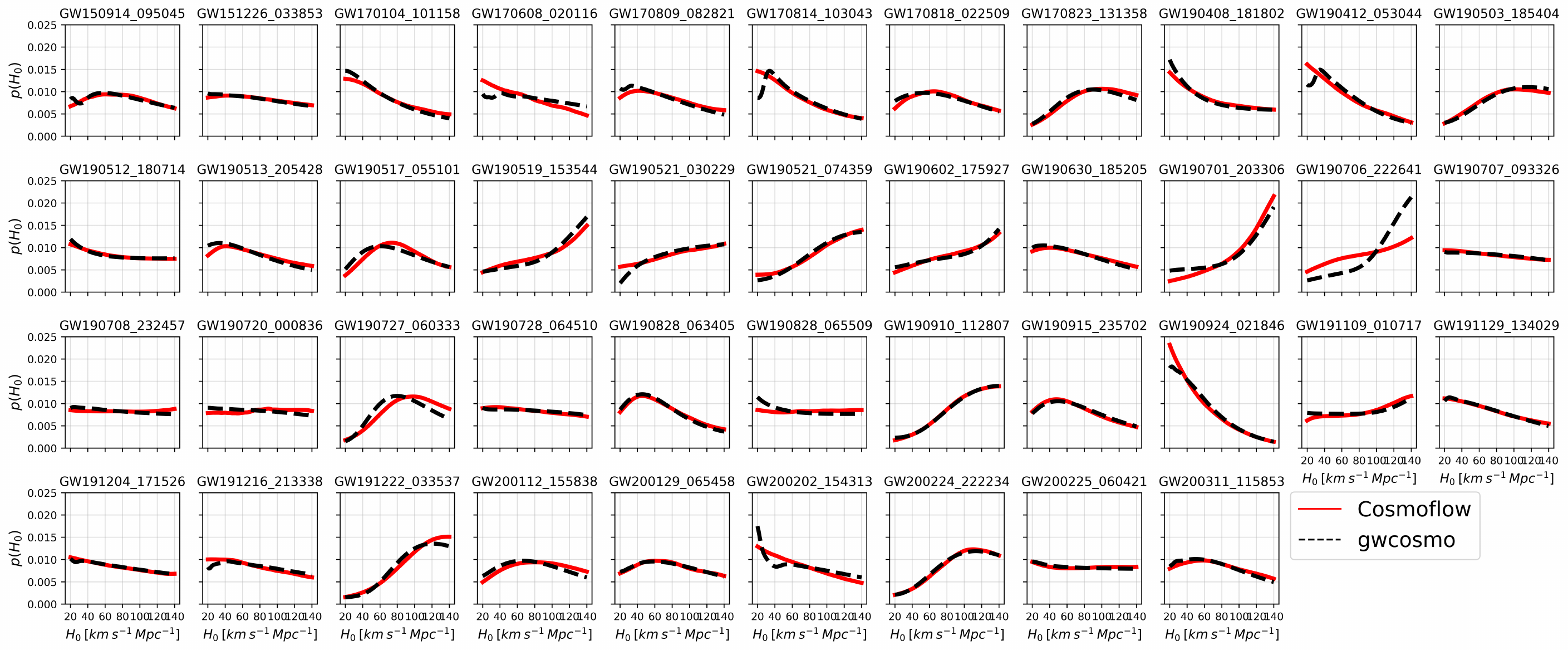}
    \caption{Likelihood distributions of $H_{0}$ for $42$ \ac{BBH} events detected during O1, O2, and O3 detection eras using \texttt{CosmoFlow} (solid red), compared with those from \cite{COSMO_2022}, using \texttt{gwcosmo} (dashed black). In all cases a uniform prior is assumed on $H_{0}$.}
    \label{fig:subplots}
\end{figure*}

We note that a direct computation of the optimal \ac{SNR} becomes a significant data generation bottleneck due to one-by-one parameter processing. By setting an \ac{SNR} threshold, many generated events are discarded, further slowing the process. Addressing this, we have developed and apply a \acf{MLP} neural network model, with a structure of 8 layers and 128 neurons per layer, to act as an accurate and efficient function approximator for the \ac{SNR} computation. Training the \ac{MLP} using 13 \ac{GW} parameters as input (omitting phase) and training with a targeted output of \ac{SNR} divided by luminosity distance, we enable efficient \ac{SNR} prediction for vectorised parameter sets, enhancing data generation speed by a factor of 20000. This \ac{MLP} implementation uses the \texttt{poplar} module~\cite{poplar} with the widely-used \texttt{pytorch} package.

\emph{Training the Normalising Flow.}---
The data used to train a \ac{NF} model for the numerator in Eq.\eqref{eq:master1} is a combination of $10^6$ samples for each observing run and detector configuration. The model considers only the luminosity distance, primary and secondary masses, and sky location, $\alpha$ and $\delta$, with the conditional parameter $\omega = H_{0}$. The remaining nine parameters are marginalized, aligning with the analysis of \cite{COSMO_2022}, which includes only \ac{GW} parameters influenced by specific cosmology or \ac{EM} catalog information.

This study uses a CouplingNF model, which uses parameterized splines to model transformations, for the best performance. Implemented using \texttt{glasflow} ~\cite{glasflow_soft}, the model has 3 block transforms, 6 layers, and 120 neurons per layer. It's trained over 500 epochs with a learning rate of 0.0005 and takes approximately 5 hours to train using a NVIDIA GeForce RTX 2080 Ti GPU.

\emph{Results.}---
We present our results in Fig.~\ref{fig:subplots}, where we show the posterior distributions over the Hubble constant for each of the \acp{BBH} observed during observing runs O1 through to O3b assuming a flat prior on $H_0$. In order to compare with results presented in~\cite{COSMO_2022} the events were selected (and the \ac{NF} models trained) based on an multi-detector \ac{SNR} threshold $\rho_{th} = 11$. We used 10,000 samples from the posterior distributions of the GWTC catalog \cite{ligo_gwtc3, url_gwtc} to evaluate the expectation value within Eq.\eqref{eq:master1}. Our individual events results are shown in comparison to the \texttt{gwcosmo} results from \cite{COSMO_2022}.

The posterior distributions over the 42 \ac{BBH} events computed using \texttt{CosmoFlow} show good agreement with the \texttt{gwcosmo} analysis results. Our analysis very effectively captures the features within the posteriors that are driven by population information, e.g., component mass distributions, co-moving volume, and merger rate. There are a handful of events whose parameters allow for a reasonable probability that they originate from a galaxy within the catalog. In these cases the \texttt{gwcosmo} posteriors exhibit structure at low $H_0$ values caused by corresponding structure within the catalog. The \texttt{CosmoFlow} analysis does not show this structure as prominently and indicates that the \ac{NF} could be developed and trained further to capture this in more detail. 

\begin{figure}
    \includegraphics[width=1.0\columnwidth]{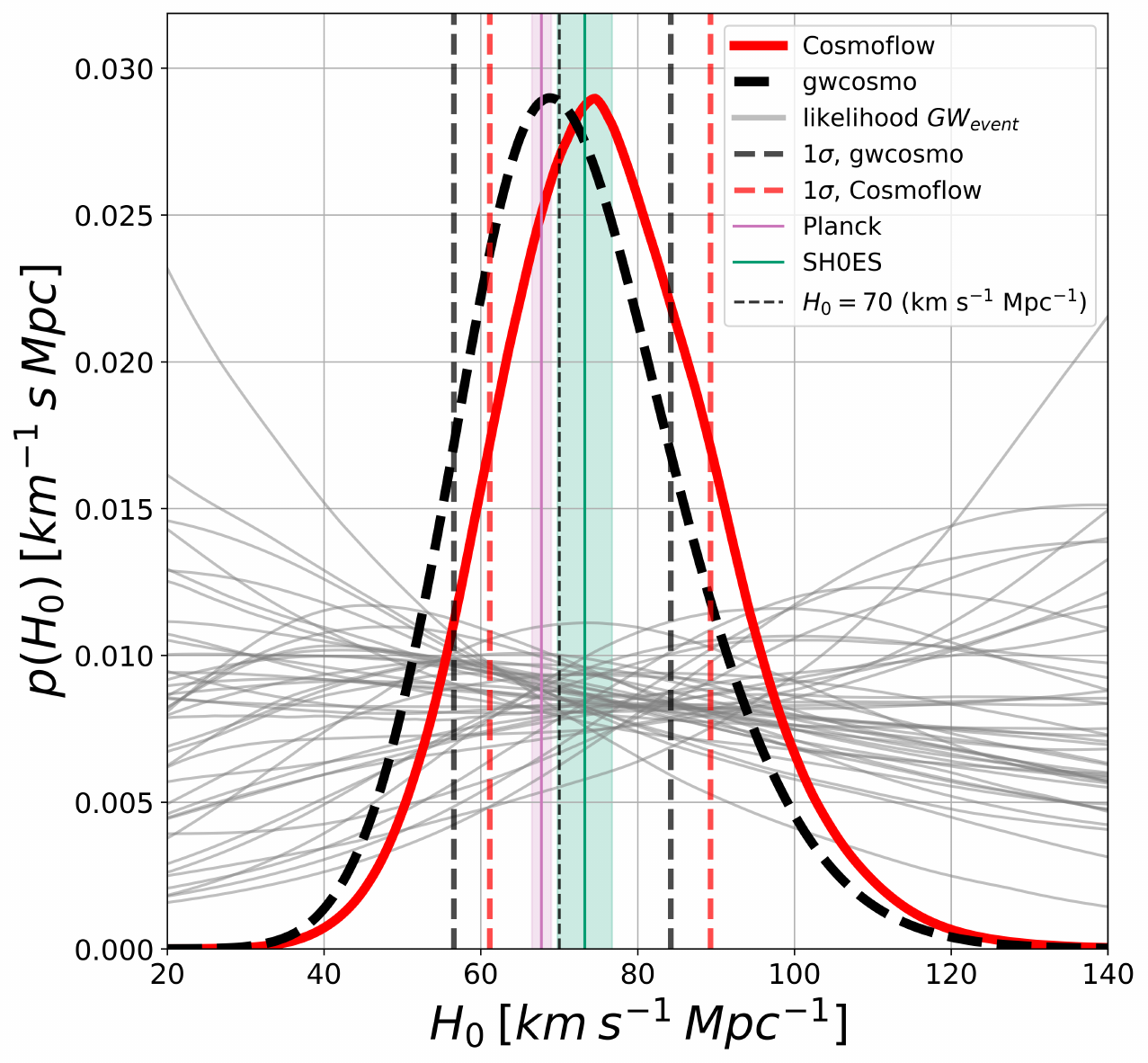}
    \caption{Combined posteriors from Fig.\ref{fig:subplots}, comparing \texttt{CosmoFlow} (solid black) with \texttt{gwcosmo} (solid blue). The single event likelihoods for \texttt{CosmoFlow} are plotted in the background in gray. Current Planck~\cite{Planck_2020} and SHOES~\cite{Riess_2019} estimates of the Hubble constant are also plotted in pink and green respectively, with 3$\sigma$ uncertainties. The 1$\sigma$ boundaries are also plotted for both \texttt{CosmoFlow} and \texttt{gwcosmo}}
    \label{fig:combined}
\end{figure}

By using Eq.~\ref{eq:master1} to combine the likelihoods from all 42 \ac{BBH} events, and assuming a flat prior, we obtain the full combined posterior distribution on $H_0$ for both methodologies. This is shown in Fig.~\ref{fig:combined}. Results show a value of  $H_{0} = 74.51^{+14.80}_{-13.63} \: \text{km} \:\text{s}^{-1} \text{Mpc}^{-1}$ ($1\sigma$ highest density interval) computed using \texttt{CosmoFlow}, compared to \texttt{gwcosmo}, which gives a measurement $H_{0} =  68.72^{+15.64}_{-12.41} \: \text{km} \:\text{s}^{-1} \text{Mpc}^{-1}$ . It is clear that the agreement between the two independent analyses is driven by the fact that both results are informed primarily through population information and the common assumptions in our models related to volumes of space outwith the galaxy catalog.

\emph{Discussion.}---
This letter presents a methodology for using a \ac{NF} via carefully generated training data to model an \ac{EM}-informed and cosmology dependent prior on \ac{GW} parameters. This fast-to-evaluate prior function can be combined with computationally-costly pre-computed posterior samples from individual \ac{GW} events to evaluate the likelihood of the \ac{GW} measurement given a specific cosmology. Results from individual events can then be used to provide a combined posterior distribution on cosmological parameters. We also provide an example analysis mirroring that of the existing standard approach where we show good agreement in our results on the posterior distribution of the Hubble constant using 42 \ac{BBH} detections from the advanced \ac{GW} detectors.

Our results exhibit good agreement with the established \texttt{gwcosmo} pipeline, a methodology employed in the advanced \ac{GW} detector O3 analysis for cosmological parameter inference. The analysis showcases strong agreement in posteriors largely informed by the population characteristics of the posterior distribution for cosmological parameter inference, as depicted by the consistent matching trends seen in Fig.~\ref{fig:subplots}. This consistency leads naturally to a strong agreement between the combined posteriors over the 42 \ac{BBH} events where the majority of inferred cosmological information stems from the population characteristics of the events. 

We note some discrepancies between our results and those of~\cite{COSMO_2022} for events that exhibit posterior structure at low values of $H_0$. These features, which are present in the \texttt{gwcosmo} posterior distributions, are not as prominent in the \texttt{CosmoFlow} posterior distributions. Reconstructing these features is challenging and demands a larger training dataset and additional development of the \ac{NF} structure to more accurately capture the galactic clustering within the GLADE+ catalog. Despite this challenge, since the discrepant posterior structure does not overlap with the area of combined posterior support (at $\sim 70\: \text{km} \:\text{s}^{-1} \text{Mpc}^{-1}$) we obtain very good agreement with the corresponding \texttt{gwcosmo} result. We also note that whilst we have used the \texttt{gwcosmo} analysis as a benchmark, it is not fair to treat it as an absolute standard since it contains (as do most computational analyses including our own) a finite number of minor approximations.
 
This method additionally introduces flexibility in selecting the number of cosmological and/or population parameters for inference as well as the number of \ac{GW} parameters to use as input. Our method requires only that training data consists of \ac{GW} parameter samples, conditional on cosmological and or population parameter samples, and that a suitably large number of these samples be generated prior to training. This enables the extension of the analysis into inference of multi-dimensional cosmological and/or population parameter spaces, where subsequent comparisons with recent improvements to \texttt{gwcosmo}~\cite{gray2023joint} can be conducted. Future work will also address the complexity of astrophysical model choices, e.g., the redshift evolution of the \ac{BBH} mass distribution parameters~\cite{Karathanasis:2022rtr}.
 
As is to be expected with \ac{GPU} powered machine learning approaches, we can also highlight the inherent computational speed of \texttt{CosmoFlow}. The evaluation of the likelihood (the expectation value in Eq.~\ref{eq:master1}) for a single conditional parameter and 1000 \ac{GW} posterior samples takes $\sim 1$ ms on a single \ac{GPU}, enabling swift computation over a range of cosmological parameter values. 

It is important to address the issue of combining joint posteriors from individual events in the case of multi-dimensional posterior distributions over cosmological and population parameters. For our 1-dimensional analysis, evaluating the likelihood for each \ac{BBH} on a common vector of $H_0$ values was sufficient. For higher dimensional cases, traditional Bayesian sampling methods can be used, e.g., Nested Sampling~\cite{Nested_sampling}, Markov chain Monte Carlo~\cite{MCMC}. However, the inherent parallelism within the \texttt{CosmoFlow} approach makes inefficient rejection sampling approaches a feasible and simple alternative. 

\emph{Acknowledgements.}---
The authors thank Michael J. Williams and Christian Chapman-Bird for insightful discussions about training normalising flows and insights for the optimisation of the data generation. The authors also thank the members of the Data Analysis Group of the Institute of Gravitational Research and the LVK Cosmology group for helpful discussions. 

F.S. is supported by the University of Glasgow CoSE. C.M. and M.H. are supported by the Science and Technology Research Council [ST/V005634/1]. This material is based upon work supported by NSF's LIGO Laboratory which is a major facility fully funded by the National Science Foundation.

\bibliography{bibliography}

\begin{thebibliography}{36}%
\makeatletter
\providecommand \@ifxundefined [1]{%
 \@ifx{#1\undefined}
}%
\providecommand \@ifnum [1]{%
 \ifnum #1\expandafter \@firstoftwo
 \else \expandafter \@secondoftwo
 \fi
}%
\providecommand \@ifx [1]{%
 \ifx #1\expandafter \@firstoftwo
 \else \expandafter \@secondoftwo
 \fi
}%
\providecommand \natexlab [1]{#1}%
\providecommand \enquote  [1]{``#1''}%
\providecommand \bibnamefont  [1]{#1}%
\providecommand \bibfnamefont [1]{#1}%
\providecommand \citenamefont [1]{#1}%
\providecommand \href@noop [0]{\@secondoftwo}%
\providecommand \href [0]{\begingroup \@sanitize@url \@href}%
\providecommand \@href[1]{\@@startlink{#1}\@@href}%
\providecommand \@@href[1]{\endgroup#1\@@endlink}%
\providecommand \@sanitize@url [0]{\catcode `\\12\catcode `\$12\catcode `\&12\catcode `\#12\catcode `\^12\catcode `\_12\catcode `\%12\relax}%
\providecommand \@@startlink[1]{}%
\providecommand \@@endlink[0]{}%
\providecommand \url  [0]{\begingroup\@sanitize@url \@url }%
\providecommand \@url [1]{\endgroup\@href {#1}{\urlprefix }}%
\providecommand \urlprefix  [0]{URL }%
\providecommand \Eprint [0]{\href }%
\providecommand \doibase [0]{https://doi.org/}%
\providecommand \selectlanguage [0]{\@gobble}%
\providecommand \bibinfo  [0]{\@secondoftwo}%
\providecommand \bibfield  [0]{\@secondoftwo}%
\providecommand \translation [1]{[#1]}%
\providecommand \BibitemOpen [0]{}%
\providecommand \bibitemStop [0]{}%
\providecommand \bibitemNoStop [0]{.\EOS\space}%
\providecommand \EOS [0]{\spacefactor3000\relax}%
\providecommand \BibitemShut  [1]{\csname bibitem#1\endcsname}%
\let\auto@bib@innerbib\@empty
\bibitem [{\citenamefont {Schutz}(1986)}]{Schutz}%
  \BibitemOpen
  \bibfield  {author} {\bibinfo {author} {\bibfnamefont {B.~F.}\ \bibnamefont {Schutz}},\ }\href@noop {} {\bibfield  {journal} {\bibinfo  {journal} {Nature}\ }\textbf {\bibinfo {volume} {323}},\ \bibinfo {pages} {310–311} (\bibinfo {year} {1986})}\BibitemShut {NoStop}%
\bibitem [{\citenamefont {{Hogg}}(1999)}]{Hogg}%
  \BibitemOpen
  \bibfield  {author} {\bibinfo {author} {\bibfnamefont {D.~W.}\ \bibnamefont {{Hogg}}},\ }\href@noop {} {\bibfield  {journal} {\bibinfo  {journal} {arXiv e-prints}\ ,\ \bibinfo {eid} {astro-ph/9905116}} (\bibinfo {year} {1999})},\ \Eprint {https://arxiv.org/abs/astro-ph/9905116} {arXiv:astro-ph/9905116 [astro-ph]} \BibitemShut {NoStop}%
\bibitem [{\citenamefont {Freedman}(2017)}]{freedman2017cosmology}%
  \BibitemOpen
  \bibfield  {author} {\bibinfo {author} {\bibfnamefont {W.~L.}\ \bibnamefont {Freedman}},\ }\href@noop {} {\bibinfo {title} {Cosmology at at crossroads: Tension with the hubble constant}} (\bibinfo {year} {2017}),\ \Eprint {https://arxiv.org/abs/1706.02739} {arXiv:1706.02739 [astro-ph.CO]} \BibitemShut {NoStop}%
\bibitem [{\citenamefont {Aghanim}\ \emph {et~al.}(2020)\citenamefont {Aghanim}, \citenamefont {Akrami}, \citenamefont {Ashdown}, \citenamefont {Aumont}, \citenamefont {Baccigalupi}, \citenamefont {Ballardini}, \citenamefont {Banday}, \citenamefont {Barreiro}, \citenamefont {Bartolo},\ and\ \citenamefont {et~al.}}]{Planck_2020}%
  \BibitemOpen
  \bibfield  {author} {\bibinfo {author} {\bibfnamefont {N.}~\bibnamefont {Aghanim}}, \bibinfo {author} {\bibfnamefont {Y.}~\bibnamefont {Akrami}}, \bibinfo {author} {\bibfnamefont {M.}~\bibnamefont {Ashdown}}, \bibinfo {author} {\bibfnamefont {J.}~\bibnamefont {Aumont}}, \bibinfo {author} {\bibfnamefont {C.}~\bibnamefont {Baccigalupi}}, \bibinfo {author} {\bibfnamefont {M.}~\bibnamefont {Ballardini}}, \bibinfo {author} {\bibfnamefont {A.~J.}\ \bibnamefont {Banday}}, \bibinfo {author} {\bibfnamefont {R.~B.}\ \bibnamefont {Barreiro}}, \bibinfo {author} {\bibfnamefont {N.}~\bibnamefont {Bartolo}},\ and\ \bibinfo {author} {\bibnamefont {et~al.}},\ }\href {https://doi.org/10.1051/0004-6361/201833910} {\bibfield  {journal} {\bibinfo  {journal} {Astronomy \& Astrophysics}\ }\textbf {\bibinfo {volume} {641}},\ \bibinfo {pages} {A6} (\bibinfo {year} {2020})}\BibitemShut {NoStop}%
\bibitem [{\citenamefont {Riess}\ \emph {et~al.}(2019)\citenamefont {Riess}, \citenamefont {Casertano}, \citenamefont {Yuan}, \citenamefont {Macri},\ and\ \citenamefont {Scolnic}}]{Riess_2019}%
  \BibitemOpen
  \bibfield  {author} {\bibinfo {author} {\bibfnamefont {A.~G.}\ \bibnamefont {Riess}}, \bibinfo {author} {\bibfnamefont {S.}~\bibnamefont {Casertano}}, \bibinfo {author} {\bibfnamefont {W.}~\bibnamefont {Yuan}}, \bibinfo {author} {\bibfnamefont {L.~M.}\ \bibnamefont {Macri}},\ and\ \bibinfo {author} {\bibfnamefont {D.}~\bibnamefont {Scolnic}},\ }\href {https://doi.org/10.3847/1538-4357/ab1422} {\bibfield  {journal} {\bibinfo  {journal} {The Astrophysical Journal}\ }\textbf {\bibinfo {volume} {876}},\ \bibinfo {pages} {85} (\bibinfo {year} {2019})}\BibitemShut {NoStop}%
\bibitem [{\citenamefont {Collabo-ration}\ and\ \citenamefont {Collaboration}(2020)}]{Abbott_2020}%
  \BibitemOpen
  \bibfield  {author} {\bibinfo {author} {\bibfnamefont {T.~L. S. C. T.~V.}\ \bibnamefont {Collabo-ration}}\ and\ \bibinfo {author} {\bibfnamefont {T.~K.}\ \bibnamefont {Collaboration}},\ }\bibfield  {journal} {\bibinfo  {journal} {Living Reviews in Relativity}\ }\textbf {\bibinfo {volume} {23}},\ \href {https://doi.org/10.1007/s41114-020-00026-9} {10.1007/s41114-020-00026-9} (\bibinfo {year} {2020})\BibitemShut {NoStop}%
\bibitem [{\citenamefont {Aasi}\ \emph {et~al.}(2015)\citenamefont {Aasi} \emph {et~al.}}]{LIGOScientific:2014pky}%
  \BibitemOpen
  \bibfield  {author} {\bibinfo {author} {\bibfnamefont {J.}~\bibnamefont {Aasi}} \emph {et~al.} (\bibinfo {collaboration} {LIGO Scientific}),\ }\href {https://doi.org/10.1088/0264-9381/32/7/074001} {\bibfield  {journal} {\bibinfo  {journal} {Class. Quant. Grav.}\ }\textbf {\bibinfo {volume} {32}},\ \bibinfo {pages} {074001} (\bibinfo {year} {2015})},\ \Eprint {https://arxiv.org/abs/1411.4547} {arXiv:1411.4547 [gr-qc]} \BibitemShut {NoStop}%
\bibitem [{\citenamefont {Acernese}\ \emph {et~al.}(2015)\citenamefont {Acernese} \emph {et~al.}}]{VIRGO:2014yos}%
  \BibitemOpen
  \bibfield  {author} {\bibinfo {author} {\bibfnamefont {F.}~\bibnamefont {Acernese}} \emph {et~al.} (\bibinfo {collaboration} {VIRGO}),\ }\href {https://doi.org/10.1088/0264-9381/32/2/024001} {\bibfield  {journal} {\bibinfo  {journal} {Class. Quant. Grav.}\ }\textbf {\bibinfo {volume} {32}},\ \bibinfo {pages} {024001} (\bibinfo {year} {2015})},\ \Eprint {https://arxiv.org/abs/1408.3978} {arXiv:1408.3978 [gr-qc]} \BibitemShut {NoStop}%
\bibitem [{\citenamefont {Collaboration}\ \emph {et~al.}(2021)\citenamefont {Collaboration}, \citenamefont {the Virgo~Collaboration},\ and\ \citenamefont {the KAGRA~Collaboration}}]{ligo_gwtc3}%
  \BibitemOpen
  \bibfield  {author} {\bibinfo {author} {\bibfnamefont {T.~L.~S.}\ \bibnamefont {Collaboration}}, \bibinfo {author} {\bibnamefont {the Virgo~Collaboration}},\ and\ \bibinfo {author} {\bibnamefont {the KAGRA~Collaboration}},\ }\href@noop {} {\bibinfo {title} {Gwtc-3: Compact binary coalescences observed by ligo and virgo during the second part of the third observing run}} (\bibinfo {year} {2021}),\ \Eprint {https://arxiv.org/abs/2111.03606} {arXiv:2111.03606 [gr-qc]} \BibitemShut {NoStop}%
\bibitem [{\citenamefont {{The LIGO Scientific Collaboration}}\ \emph {et~al.}(2021{\natexlab{a}})\citenamefont {{The LIGO Scientific Collaboration}}, \citenamefont {{The Virgo Collaboration}},\ and\ \citenamefont {{The KAGRA Collaboration}}}]{Pop_paper_2021}%
  \BibitemOpen
  \bibfield  {author} {\bibinfo {author} {\bibnamefont {{The LIGO Scientific Collaboration}}}, \bibinfo {author} {\bibnamefont {{The Virgo Collaboration}}},\ and\ \bibinfo {author} {\bibnamefont {{The KAGRA Collaboration}}},\ }\href {https://doi.org/10.48550/ARXIV.2111.03634} {\bibinfo {title} {The population of merging compact binaries inferred using gravitational waves through gwtc-3}} (\bibinfo {year} {2021}{\natexlab{a}})\BibitemShut {NoStop}%
\bibitem [{\citenamefont {{The LIGO Scientific Collaboration}}\ \emph {et~al.}(2021{\natexlab{b}})\citenamefont {{The LIGO Scientific Collaboration}}, \citenamefont {{The Virgo Collaboration}},\ and\ \citenamefont {{The KAGRA Collaboration}}}]{COSMO_2022}%
  \BibitemOpen
  \bibfield  {author} {\bibinfo {author} {\bibnamefont {{The LIGO Scientific Collaboration}}}, \bibinfo {author} {\bibnamefont {{The Virgo Collaboration}}},\ and\ \bibinfo {author} {\bibnamefont {{The KAGRA Collaboration}}},\ }\href {https://doi.org/10.48550/ARXIV.2111.03604} {\bibinfo {title} {Constraints on the cosmic expansion history from gwtc-3}} (\bibinfo {year} {2021}{\natexlab{b}})\BibitemShut {NoStop}%
\bibitem [{\citenamefont {Gray}\ \emph {et~al.}(2020{\natexlab{a}})\citenamefont {Gray}, \citenamefont {Hernandez}, \citenamefont {Qi}, \citenamefont {Sur}, \citenamefont {Brady}, \citenamefont {Chen}, \citenamefont {Farr}, \citenamefont {Fishbach}, \citenamefont {Gair}, \citenamefont {Ghosh}, \citenamefont {Holz}, \citenamefont {Mastrogiovanni}, \citenamefont {Messenger}, \citenamefont {Steer},\ and\ \citenamefont {Veitch}}]{PhysRevD.101.122001}%
  \BibitemOpen
  \bibfield  {author} {\bibinfo {author} {\bibfnamefont {R.}~\bibnamefont {Gray}}, \bibinfo {author} {\bibfnamefont {I.~M.~n.}\ \bibnamefont {Hernandez}}, \bibinfo {author} {\bibfnamefont {H.}~\bibnamefont {Qi}}, \bibinfo {author} {\bibfnamefont {A.}~\bibnamefont {Sur}}, \bibinfo {author} {\bibfnamefont {P.~R.}\ \bibnamefont {Brady}}, \bibinfo {author} {\bibfnamefont {H.-Y.}\ \bibnamefont {Chen}}, \bibinfo {author} {\bibfnamefont {W.~M.}\ \bibnamefont {Farr}}, \bibinfo {author} {\bibfnamefont {M.}~\bibnamefont {Fishbach}}, \bibinfo {author} {\bibfnamefont {J.~R.}\ \bibnamefont {Gair}}, \bibinfo {author} {\bibfnamefont {A.}~\bibnamefont {Ghosh}}, \bibinfo {author} {\bibfnamefont {D.~E.}\ \bibnamefont {Holz}}, \bibinfo {author} {\bibfnamefont {S.}~\bibnamefont {Mastrogiovanni}}, \bibinfo {author} {\bibfnamefont {C.}~\bibnamefont {Messenger}}, \bibinfo {author} {\bibfnamefont {D.~A.}\ \bibnamefont {Steer}},\ and\ \bibinfo {author} {\bibfnamefont {J.}~\bibnamefont {Veitch}},\ }\href
  {https://doi.org/10.1103/PhysRevD.101.122001} {\bibfield  {journal} {\bibinfo  {journal} {Phys. Rev. D}\ }\textbf {\bibinfo {volume} {101}},\ \bibinfo {pages} {122001} (\bibinfo {year} {2020}{\natexlab{a}})}\BibitemShut {NoStop}%
\bibitem [{\citenamefont {Mastrogiovanni}\ \emph {et~al.}(2023)\citenamefont {Mastrogiovanni}, \citenamefont {Pierra}, \citenamefont {Perriès}, \citenamefont {Laghi}, \citenamefont {Santoro}, \citenamefont {Ghosh}, \citenamefont {Gray}, \citenamefont {Karathanasis},\ and\ \citenamefont {Leyde}}]{ICAROGW}%
  \BibitemOpen
  \bibfield  {author} {\bibinfo {author} {\bibfnamefont {S.}~\bibnamefont {Mastrogiovanni}}, \bibinfo {author} {\bibfnamefont {G.}~\bibnamefont {Pierra}}, \bibinfo {author} {\bibfnamefont {S.}~\bibnamefont {Perriès}}, \bibinfo {author} {\bibfnamefont {D.}~\bibnamefont {Laghi}}, \bibinfo {author} {\bibfnamefont {G.~C.}\ \bibnamefont {Santoro}}, \bibinfo {author} {\bibfnamefont {A.}~\bibnamefont {Ghosh}}, \bibinfo {author} {\bibfnamefont {R.}~\bibnamefont {Gray}}, \bibinfo {author} {\bibfnamefont {C.}~\bibnamefont {Karathanasis}},\ and\ \bibinfo {author} {\bibfnamefont {K.}~\bibnamefont {Leyde}},\ }\href@noop {} {\bibinfo {title} {Icarogw: A python package for inference of astrophysical population properties of noisy, heterogeneous and incomplete observations}} (\bibinfo {year} {2023}),\ \Eprint {https://arxiv.org/abs/2305.17973} {arXiv:2305.17973 [astro-ph.CO]} \BibitemShut {NoStop}%
\bibitem [{\citenamefont {Mukherjee}\ \emph {et~al.}(2022)\citenamefont {Mukherjee}, \citenamefont {Krolewski}, \citenamefont {Wandelt},\ and\ \citenamefont {Silk}}]{Mukherjee:2022afz}%
  \BibitemOpen
  \bibfield  {author} {\bibinfo {author} {\bibfnamefont {S.}~\bibnamefont {Mukherjee}}, \bibinfo {author} {\bibfnamefont {A.}~\bibnamefont {Krolewski}}, \bibinfo {author} {\bibfnamefont {B.~D.}\ \bibnamefont {Wandelt}},\ and\ \bibinfo {author} {\bibfnamefont {J.}~\bibnamefont {Silk}},\ }\href@noop {} {\bibfield  {journal} {\bibinfo  {journal} {arXiv}\ } (\bibinfo {year} {2022})},\ \Eprint {https://arxiv.org/abs/2203.03643} {arXiv:2203.03643 [astro-ph.CO]} \BibitemShut {NoStop}%
\bibitem [{\citenamefont {Mukherjee}\ \emph {et~al.}(2021)\citenamefont {Mukherjee}, \citenamefont {Wandelt}, \citenamefont {Nissanke},\ and\ \citenamefont {Silvestri}}]{Mukherjee:2020hyn}%
  \BibitemOpen
  \bibfield  {author} {\bibinfo {author} {\bibfnamefont {S.}~\bibnamefont {Mukherjee}}, \bibinfo {author} {\bibfnamefont {B.~D.}\ \bibnamefont {Wandelt}}, \bibinfo {author} {\bibfnamefont {S.~M.}\ \bibnamefont {Nissanke}},\ and\ \bibinfo {author} {\bibfnamefont {A.}~\bibnamefont {Silvestri}},\ }\href {https://doi.org/10.1103/PhysRevD.103.043520} {\bibfield  {journal} {\bibinfo  {journal} {Phys. Rev. D}\ }\textbf {\bibinfo {volume} {103}},\ \bibinfo {pages} {043520} (\bibinfo {year} {2021})},\ \Eprint {https://arxiv.org/abs/2007.02943} {arXiv:2007.02943 [astro-ph.CO]} \BibitemShut {NoStop}%
\bibitem [{\citenamefont {Diaz}\ and\ \citenamefont {Mukherjee}(2022)}]{Diaz:2021pem}%
  \BibitemOpen
  \bibfield  {author} {\bibinfo {author} {\bibfnamefont {C.~C.}\ \bibnamefont {Diaz}}\ and\ \bibinfo {author} {\bibfnamefont {S.}~\bibnamefont {Mukherjee}},\ }\href {https://doi.org/10.1093/mnras/stac208} {\bibfield  {journal} {\bibinfo  {journal} {Mon. Not. Roy. Astron. Soc.}\ }\textbf {\bibinfo {volume} {511}},\ \bibinfo {pages} {2782} (\bibinfo {year} {2022})},\ \Eprint {https://arxiv.org/abs/2107.12787} {2107.12787 [astro-ph.CO]} \BibitemShut {NoStop}%
\bibitem [{\citenamefont {{Kobyzev}}\ \emph {et~al.}(2020)\citenamefont {{Kobyzev}}, \citenamefont {{Prince}},\ and\ \citenamefont {{Brubaker}}}]{NF}%
  \BibitemOpen
  \bibfield  {author} {\bibinfo {author} {\bibfnamefont {I.}~\bibnamefont {{Kobyzev}}}, \bibinfo {author} {\bibfnamefont {S.}~\bibnamefont {{Prince}}},\ and\ \bibinfo {author} {\bibfnamefont {M.}~\bibnamefont {{Brubaker}}},\ }\href {https://doi.org/10.1109/TPAMI.2020.2992934} {\bibfield  {journal} {\bibinfo  {journal} {IEEE Transactions on Pattern Analysis and Machine Intelligence}\ ,\ \bibinfo {pages} {1}} (\bibinfo {year} {2020})}\BibitemShut {NoStop}%
\bibitem [{\citenamefont {Williams}\ \emph {et~al.}(2021)\citenamefont {Williams}, \citenamefont {Veitch},\ and\ \citenamefont {Messenger}}]{Williams_2021}%
  \BibitemOpen
  \bibfield  {author} {\bibinfo {author} {\bibfnamefont {M.~J.}\ \bibnamefont {Williams}}, \bibinfo {author} {\bibfnamefont {J.}~\bibnamefont {Veitch}},\ and\ \bibinfo {author} {\bibfnamefont {C.}~\bibnamefont {Messenger}},\ }\bibfield  {journal} {\bibinfo  {journal} {Physical Review D}\ }\textbf {\bibinfo {volume} {103}},\ \href {https://doi.org/10.1103/physrevd.103.103006} {10.1103/physrevd.103.103006} (\bibinfo {year} {2021})\BibitemShut {NoStop}%
\bibitem [{\citenamefont {Williams}\ \emph {et~al.}(2023{\natexlab{a}})\citenamefont {Williams}, \citenamefont {Veitch},\ and\ \citenamefont {Messenger}}]{williams2023importance}%
  \BibitemOpen
  \bibfield  {author} {\bibinfo {author} {\bibfnamefont {M.~J.}\ \bibnamefont {Williams}}, \bibinfo {author} {\bibfnamefont {J.}~\bibnamefont {Veitch}},\ and\ \bibinfo {author} {\bibfnamefont {C.}~\bibnamefont {Messenger}},\ }\href@noop {} {\bibinfo {title} {Importance nested sampling with normalising flows}} (\bibinfo {year} {2023}{\natexlab{a}}),\ \Eprint {https://arxiv.org/abs/2302.08526} {arXiv:2302.08526 [astro-ph.IM]} \BibitemShut {NoStop}%
\bibitem [{\citenamefont {Dax}\ \emph {et~al.}(2021)\citenamefont {Dax}, \citenamefont {Green}, \citenamefont {Gair}, \citenamefont {Macke}, \citenamefont {Buonanno},\ and\ \citenamefont {Schölkopf}}]{DINGO}%
  \BibitemOpen
  \bibfield  {author} {\bibinfo {author} {\bibfnamefont {M.}~\bibnamefont {Dax}}, \bibinfo {author} {\bibfnamefont {S.~R.}\ \bibnamefont {Green}}, \bibinfo {author} {\bibfnamefont {J.}~\bibnamefont {Gair}}, \bibinfo {author} {\bibfnamefont {J.~H.}\ \bibnamefont {Macke}}, \bibinfo {author} {\bibfnamefont {A.}~\bibnamefont {Buonanno}},\ and\ \bibinfo {author} {\bibfnamefont {B.}~\bibnamefont {Schölkopf}},\ }\bibfield  {journal} {\bibinfo  {journal} {Physical Review Letters}\ }\textbf {\bibinfo {volume} {127}},\ \href {https://doi.org/10.1103/physrevlett.127.241103} {10.1103/physrevlett.127.241103} (\bibinfo {year} {2021})\BibitemShut {NoStop}%
\bibitem [{\citenamefont {Ruhe}\ \emph {et~al.}(2022)\citenamefont {Ruhe}, \citenamefont {Wong}, \citenamefont {Cranmer},\ and\ \citenamefont {Forré}}]{ruhe2022normalizing}%
  \BibitemOpen
  \bibfield  {author} {\bibinfo {author} {\bibfnamefont {D.}~\bibnamefont {Ruhe}}, \bibinfo {author} {\bibfnamefont {K.}~\bibnamefont {Wong}}, \bibinfo {author} {\bibfnamefont {M.}~\bibnamefont {Cranmer}},\ and\ \bibinfo {author} {\bibfnamefont {P.}~\bibnamefont {Forré}},\ }\href@noop {} {\bibinfo {title} {Normalizing flows for hierarchical bayesian analysis: A gravitational wave population study}} (\bibinfo {year} {2022}),\ \Eprint {https://arxiv.org/abs/2211.09008} {arXiv:2211.09008 [astro-ph.IM]} \BibitemShut {NoStop}%
\bibitem [{\citenamefont {Papamakarios}\ \emph {et~al.}(2019)\citenamefont {Papamakarios}, \citenamefont {Nalisnick}, \citenamefont {Rezende}, \citenamefont {Mohamed},\ and\ \citenamefont {Lakshminarayanan}}]{NF_papa}%
  \BibitemOpen
  \bibfield  {author} {\bibinfo {author} {\bibfnamefont {G.}~\bibnamefont {Papamakarios}}, \bibinfo {author} {\bibfnamefont {E.}~\bibnamefont {Nalisnick}}, \bibinfo {author} {\bibfnamefont {D.~J.}\ \bibnamefont {Rezende}}, \bibinfo {author} {\bibfnamefont {S.}~\bibnamefont {Mohamed}},\ and\ \bibinfo {author} {\bibfnamefont {B.}~\bibnamefont {Lakshminarayanan}},\ }\bibfield  {journal} {\bibinfo  {journal} {Journal of Machine Learning Research}\ }\href {https://doi.org/10.48550/ARXIV.1912.02762} {10.48550/ARXIV.1912.02762} (\bibinfo {year} {2019})\BibitemShut {NoStop}%
\bibitem [{\citenamefont {Papamakarios}\ \emph {et~al.}(2018)\citenamefont {Papamakarios}, \citenamefont {Pavlakou},\ and\ \citenamefont {Murray}}]{papamakarios2018masked}%
  \BibitemOpen
  \bibfield  {author} {\bibinfo {author} {\bibfnamefont {G.}~\bibnamefont {Papamakarios}}, \bibinfo {author} {\bibfnamefont {T.}~\bibnamefont {Pavlakou}},\ and\ \bibinfo {author} {\bibfnamefont {I.}~\bibnamefont {Murray}},\ }\href@noop {} {\bibinfo {title} {Masked autoregressive flow for density estimation}} (\bibinfo {year} {2018}),\ \Eprint {https://arxiv.org/abs/1705.07057} {arXiv:1705.07057 [stat.ML]} \BibitemShut {NoStop}%
\bibitem [{\citenamefont {Winkler}\ \emph {et~al.}(2019)\citenamefont {Winkler}, \citenamefont {Worrall}, \citenamefont {Hoogeboom},\ and\ \citenamefont {Welling}}]{conditionalflow}%
  \BibitemOpen
  \bibfield  {author} {\bibinfo {author} {\bibfnamefont {C.}~\bibnamefont {Winkler}}, \bibinfo {author} {\bibfnamefont {D.}~\bibnamefont {Worrall}}, \bibinfo {author} {\bibfnamefont {E.}~\bibnamefont {Hoogeboom}},\ and\ \bibinfo {author} {\bibfnamefont {M.}~\bibnamefont {Welling}},\ }\href@noop {} {\bibinfo {title} {Learning likelihoods with conditional normalizing flows}} (\bibinfo {year} {2019}),\ \Eprint {https://arxiv.org/abs/1912.00042} {arXiv:1912.00042 [cs.LG]} \BibitemShut {NoStop}%
\bibitem [{\citenamefont {Gray}\ \emph {et~al.}(2020{\natexlab{b}})\citenamefont {Gray}, \citenamefont {Hernandez}, \citenamefont {Qi}, \citenamefont {Sur}, \citenamefont {Brady}, \citenamefont {Chen}, \citenamefont {Farr}, \citenamefont {Fishbach}, \citenamefont {Gair}, \citenamefont {Ghosh},\ and\ \citenamefont {et~al.}}]{Gray_2020}%
  \BibitemOpen
  \bibfield  {author} {\bibinfo {author} {\bibfnamefont {R.}~\bibnamefont {Gray}}, \bibinfo {author} {\bibfnamefont {I.~M.}\ \bibnamefont {Hernandez}}, \bibinfo {author} {\bibfnamefont {H.}~\bibnamefont {Qi}}, \bibinfo {author} {\bibfnamefont {A.}~\bibnamefont {Sur}}, \bibinfo {author} {\bibfnamefont {P.~R.}\ \bibnamefont {Brady}}, \bibinfo {author} {\bibfnamefont {H.-Y.}\ \bibnamefont {Chen}}, \bibinfo {author} {\bibfnamefont {W.~M.}\ \bibnamefont {Farr}}, \bibinfo {author} {\bibfnamefont {M.}~\bibnamefont {Fishbach}}, \bibinfo {author} {\bibfnamefont {J.~R.}\ \bibnamefont {Gair}}, \bibinfo {author} {\bibfnamefont {A.}~\bibnamefont {Ghosh}},\ and\ \bibinfo {author} {\bibnamefont {et~al.}},\ }\bibfield  {journal} {\bibinfo  {journal} {Physical Review D}\ }\textbf {\bibinfo {volume} {101}},\ \href {https://doi.org/10.1103/physrevd.101.122001} {10.1103/physrevd.101.122001} (\bibinfo {year} {2020}{\natexlab{b}})\BibitemShut {NoStop}%
\bibitem [{\citenamefont {D{\'{a} }lya}\ \emph {et~al.}(2018)\citenamefont {D{\'{a} }lya}, \citenamefont {Galg{\'{o}}czi}, \citenamefont {Dobos}, \citenamefont {Frei}, \citenamefont {Heng}, \citenamefont {Macas}, \citenamefont {Messenger}, \citenamefont {Raffai},\ and\ \citenamefont {de~Souza}}]{GLADE}%
  \BibitemOpen
  \bibfield  {author} {\bibinfo {author} {\bibfnamefont {G.}~\bibnamefont {D{\'{a} }lya}}, \bibinfo {author} {\bibfnamefont {G.}~\bibnamefont {Galg{\'{o}}czi}}, \bibinfo {author} {\bibfnamefont {L.}~\bibnamefont {Dobos}}, \bibinfo {author} {\bibfnamefont {Z.}~\bibnamefont {Frei}}, \bibinfo {author} {\bibfnamefont {I.~S.}\ \bibnamefont {Heng}}, \bibinfo {author} {\bibfnamefont {R.}~\bibnamefont {Macas}}, \bibinfo {author} {\bibfnamefont {C.}~\bibnamefont {Messenger}}, \bibinfo {author} {\bibfnamefont {P.}~\bibnamefont {Raffai}},\ and\ \bibinfo {author} {\bibfnamefont {R.~S.}\ \bibnamefont {de~Souza}},\ }\href {https://doi.org/10.1093/mnras/sty1703} {\bibfield  {journal} {\bibinfo  {journal} {Monthly Notices of the Royal Astronomical Society}\ }\textbf {\bibinfo {volume} {479}},\ \bibinfo {pages} {2374} (\bibinfo {year} {2018})}\BibitemShut {NoStop}%
\bibitem [{Note1()}]{Note1}%
  \BibitemOpen
  \bibinfo {note} {In general this distribution is dependent on the cosmological parameters $\Omega $ with the exception of the Hubble constant.}\BibitemShut {Stop}%
\bibitem [{\citenamefont {Gray}(2021)}]{GrayTh}%
  \BibitemOpen
  \bibfield  {author} {\bibinfo {author} {\bibfnamefont {R.}~\bibnamefont {Gray}},\ }\href@noop {} {\bibfield  {journal} {\bibinfo  {journal} {https://theses.gla.ac.uk/82438/}\ } (\bibinfo {year} {2021})}\BibitemShut {NoStop}%
\bibitem [{\citenamefont {Ashton}\ \emph {et~al.}(2019)\citenamefont {Ashton}, \citenamefont {Hübner}, \citenamefont {Lasky}, \citenamefont {Talbot}, \citenamefont {Ackley}, \citenamefont {Biscoveanu}, \citenamefont {Chu}, \citenamefont {Divakarla}, \citenamefont {Easter}, \citenamefont {Goncharov}, \citenamefont {Vivanco}, \citenamefont {Harms}, \citenamefont {Lower}, \citenamefont {Meadors}, \citenamefont {Melchor}, \citenamefont {Payne}, \citenamefont {Pitkin}, \citenamefont {Powell}, \citenamefont {Sarin}, \citenamefont {Smith},\ and\ \citenamefont {Thrane}}]{Bilby}%
  \BibitemOpen
  \bibfield  {author} {\bibinfo {author} {\bibfnamefont {G.}~\bibnamefont {Ashton}}, \bibinfo {author} {\bibfnamefont {M.}~\bibnamefont {Hübner}}, \bibinfo {author} {\bibfnamefont {P.~D.}\ \bibnamefont {Lasky}}, \bibinfo {author} {\bibfnamefont {C.}~\bibnamefont {Talbot}}, \bibinfo {author} {\bibfnamefont {K.}~\bibnamefont {Ackley}}, \bibinfo {author} {\bibfnamefont {S.}~\bibnamefont {Biscoveanu}}, \bibinfo {author} {\bibfnamefont {Q.}~\bibnamefont {Chu}}, \bibinfo {author} {\bibfnamefont {A.}~\bibnamefont {Divakarla}}, \bibinfo {author} {\bibfnamefont {P.~J.}\ \bibnamefont {Easter}}, \bibinfo {author} {\bibfnamefont {B.}~\bibnamefont {Goncharov}}, \bibinfo {author} {\bibfnamefont {F.~H.}\ \bibnamefont {Vivanco}}, \bibinfo {author} {\bibfnamefont {J.}~\bibnamefont {Harms}}, \bibinfo {author} {\bibfnamefont {M.~E.}\ \bibnamefont {Lower}}, \bibinfo {author} {\bibfnamefont {G.~D.}\ \bibnamefont {Meadors}}, \bibinfo {author} {\bibfnamefont {D.}~\bibnamefont {Melchor}}, \bibinfo {author} {\bibfnamefont
  {E.}~\bibnamefont {Payne}}, \bibinfo {author} {\bibfnamefont {M.~D.}\ \bibnamefont {Pitkin}}, \bibinfo {author} {\bibfnamefont {J.}~\bibnamefont {Powell}}, \bibinfo {author} {\bibfnamefont {N.}~\bibnamefont {Sarin}}, \bibinfo {author} {\bibfnamefont {R.~J.~E.}\ \bibnamefont {Smith}},\ and\ \bibinfo {author} {\bibfnamefont {E.}~\bibnamefont {Thrane}},\ }\href {https://doi.org/10.3847/1538-4365/ab06fc} {\bibfield  {journal} {\bibinfo  {journal} {The Astrophysical Journal Supplement Series}\ }\textbf {\bibinfo {volume} {241}},\ \bibinfo {pages} {27} (\bibinfo {year} {2019})}\BibitemShut {NoStop}%
\bibitem [{\citenamefont {Chapman-Bird}\ \emph {et~al.}(2023)\citenamefont {Chapman-Bird}, \citenamefont {Berry},\ and\ \citenamefont {Woan}}]{poplar}%
  \BibitemOpen
  \bibfield  {author} {\bibinfo {author} {\bibfnamefont {C.~E.~A.}\ \bibnamefont {Chapman-Bird}}, \bibinfo {author} {\bibfnamefont {C.~P.~L.}\ \bibnamefont {Berry}},\ and\ \bibinfo {author} {\bibfnamefont {G.}~\bibnamefont {Woan}},\ }\href {https://doi.org/10.1093/mnras/stad1397} {\bibfield  {journal} {\bibinfo  {journal} {Monthly Notices of the Royal Astronomical Society}\ }\textbf {\bibinfo {volume} {522}},\ \bibinfo {pages} {6043} (\bibinfo {year} {2023})},\ \Eprint {https://arxiv.org/abs/https://academic.oup.com/mnras/article-pdf/522/4/6043/50390741/stad1397.pdf} {https://academic.oup.com/mnras/article-pdf/522/4/6043/50390741/stad1397.pdf} \BibitemShut {NoStop}%
\bibitem [{\citenamefont {Williams}\ \emph {et~al.}(2023{\natexlab{b}})\citenamefont {Williams}, \citenamefont {Mcginn}, \citenamefont {Stachurski},\ and\ \citenamefont {Veitch}}]{glasflow_soft}%
  \BibitemOpen
  \bibfield  {author} {\bibinfo {author} {\bibfnamefont {M.~J.}\ \bibnamefont {Williams}}, \bibinfo {author} {\bibfnamefont {J.}~\bibnamefont {Mcginn}}, \bibinfo {author} {\bibfnamefont {F.}~\bibnamefont {Stachurski}},\ and\ \bibinfo {author} {\bibfnamefont {J.}~\bibnamefont {Veitch}},\ }\href {https://doi.org/10.5281/zenodo.7598678} {\bibinfo {title} {uofgravity/glasflow: v0.2.0}} (\bibinfo {year} {2023}{\natexlab{b}})\BibitemShut {NoStop}%
\bibitem [{url(2021)}]{url_gwtc}%
  \BibitemOpen
  \href@noop {} {}\bibinfo {howpublished} {\url{https://dcc.ligo.org/LIGO-P2000318/public}} (\bibinfo {year} {2021})\BibitemShut {NoStop}%
\bibitem [{\citenamefont {Gray}\ \emph {et~al.}(2023)\citenamefont {Gray}, \citenamefont {Beirnaert}, \citenamefont {Karathanasis}, \citenamefont {Revenu}, \citenamefont {Turski}, \citenamefont {Chen}, \citenamefont {Baker}, \citenamefont {Vallejo}, \citenamefont {Romano}, \citenamefont {Ghosh}, \citenamefont {Ghosh}, \citenamefont {Leyde}, \citenamefont {Mastrogiovanni},\ and\ \citenamefont {More}}]{gray2023joint}%
  \BibitemOpen
  \bibfield  {author} {\bibinfo {author} {\bibfnamefont {R.}~\bibnamefont {Gray}}, \bibinfo {author} {\bibfnamefont {F.}~\bibnamefont {Beirnaert}}, \bibinfo {author} {\bibfnamefont {C.}~\bibnamefont {Karathanasis}}, \bibinfo {author} {\bibfnamefont {B.}~\bibnamefont {Revenu}}, \bibinfo {author} {\bibfnamefont {C.}~\bibnamefont {Turski}}, \bibinfo {author} {\bibfnamefont {A.}~\bibnamefont {Chen}}, \bibinfo {author} {\bibfnamefont {T.}~\bibnamefont {Baker}}, \bibinfo {author} {\bibfnamefont {S.}~\bibnamefont {Vallejo}}, \bibinfo {author} {\bibfnamefont {A.~E.}\ \bibnamefont {Romano}}, \bibinfo {author} {\bibfnamefont {T.}~\bibnamefont {Ghosh}}, \bibinfo {author} {\bibfnamefont {A.}~\bibnamefont {Ghosh}}, \bibinfo {author} {\bibfnamefont {K.}~\bibnamefont {Leyde}}, \bibinfo {author} {\bibfnamefont {S.}~\bibnamefont {Mastrogiovanni}},\ and\ \bibinfo {author} {\bibfnamefont {S.}~\bibnamefont {More}},\ }\href@noop {} {\bibinfo {title} {Joint cosmological and gravitational-wave population inference using dark sirens
  and galaxy catalogues}} (\bibinfo {year} {2023}),\ \Eprint {https://arxiv.org/abs/2308.02281} {arXiv:2308.02281 [astro-ph.CO]} \BibitemShut {NoStop}%
\bibitem [{\citenamefont {Karathanasis}\ \emph {et~al.}(2023)\citenamefont {Karathanasis}, \citenamefont {Mukherjee},\ and\ \citenamefont {Mastrogiovanni}}]{Karathanasis:2022rtr}%
  \BibitemOpen
  \bibfield  {author} {\bibinfo {author} {\bibfnamefont {C.}~\bibnamefont {Karathanasis}}, \bibinfo {author} {\bibfnamefont {S.}~\bibnamefont {Mukherjee}},\ and\ \bibinfo {author} {\bibfnamefont {S.}~\bibnamefont {Mastrogiovanni}},\ }\href {https://doi.org/10.1093/mnras/stad1373} {\bibfield  {journal} {\bibinfo  {journal} {Mon. Not. Roy. Astron. Soc.}\ }\textbf {\bibinfo {volume} {523}},\ \bibinfo {pages} {4539} (\bibinfo {year} {2023})},\ \Eprint {https://arxiv.org/abs/2204.13495} {arXiv:2204.13495 [astro-ph.CO]} \BibitemShut {NoStop}%
\bibitem [{\citenamefont {Skilling}(2004)}]{Nested_sampling}%
  \BibitemOpen
  \bibfield  {author} {\bibinfo {author} {\bibfnamefont {J.}~\bibnamefont {Skilling}},\ }\href {https://doi.org/10.1063/1.1835238} {\bibfield  {journal} {\bibinfo  {journal} {AIP Conference Proceedings}\ }\textbf {\bibinfo {volume} {735}},\ \bibinfo {pages} {395} (\bibinfo {year} {2004})},\ \Eprint {https://arxiv.org/abs/https://pubs.aip.org/aip/acp/article-pdf/735/1/395/11702789/395\_1\_online.pdf} {https://pubs.aip.org/aip/acp/article-pdf/735/1/395/11702789/395\_1\_online.pdf} \BibitemShut {NoStop}%
\bibitem [{\citenamefont {Hastings}(1970)}]{MCMC}%
  \BibitemOpen
  \bibfield  {author} {\bibinfo {author} {\bibfnamefont {W.~K.}\ \bibnamefont {Hastings}},\ }\href {https://doi.org/10.1093/biomet/57.1.97} {\bibfield  {journal} {\bibinfo  {journal} {Biometrika}\ }\textbf {\bibinfo {volume} {57}},\ \bibinfo {pages} {97} (\bibinfo {year} {1970})},\ \Eprint {https://arxiv.org/abs/https://academic.oup.com/biomet/article-pdf/57/1/97/23940249/57-1-97.pdf} {https://academic.oup.com/biomet/article-pdf/57/1/97/23940249/57-1-97.pdf} \BibitemShut {NoStop}%
\end{thebibliography}%
\end{document}